%   Compilare in PlainTeX

\mag=1300

%%%%%%%%%%%%%%%%%%%%%%%%  Font %%%%%%%%%%%%%%%%%%%%%%%%%%%%%%

\font\eightrm=cmr8
\font\sevenrm=cmr7
\font\seventeenbf=cmbx12 at 14pt

\font\tenrm=cmr10
\font\tenit=cmti10
\font\tensl=cmsl10
\font\tenbf=cmbx10
\font\tentt=cmtt10

\def\tenpoint{%
\def\rm{\fam0\tenrm}%
\def\it{\fam\itfam\tenit}%
\def\sl{\fam\slfam\tensl}%
\def\tt{\fam\ttfam\tentt}%
\def\bf{\fam\bffam\tenbf}%
}

\tenpoint\rm

%%%%%%%%%%%%%%%%%%%%%%%%         %%%%%%%%%%%%%%%%%%%%%%%%%%%%%%
%%%%%%%%%%%%%%%%%%%%%%%%  Format %%%%%%%%%%%%%%%%%%%%%%%%%%%%%%
\newcount\notenumber

\def\note{\global\advance\notenumber by 1
\footnote{{\mathsurround=0pt$^{\the\notenumber}$}}
}

\normalbaselineskip=11.66pt
\normallineskip=2pt minus 1pt
\normallineskiplimit=1pt
\normalbaselines
\vsize=17.3cm
\hsize=12.1cm
\parindent=20pt
\smallskipamount=3.6pt plus 1pt minus 9pt
\abovedisplayskip=1\normalbaselineskip plus 3pt minus 9pt
\belowdisplayskip=1\normalbaselineskip plus 3pt minus 9pt
\skip\footins=2\baselineskip
\advance\skip\footins by 3pt

\mathsurround 2pt
\newdimen\leftind \leftind=0cm
\newdimen\rightind \rightind=0.65cm

\def\pagenumbers{\footline={\hss\tenrm\folio\hss}}
\nopagenumbers

\def\MainHead{{\baselineskip 16.7pt\seventeenbf
\noindent\titolo\par}
\normalbaselines
%\vskip 28.34pt
\vskip 18.34pt
\noindent\autori\par
\ni\indirizzo
%\footnote{\phantom{i}}{\piedipagina}
%\vskip 73.34pt
\vskip 23.34pt
\ni {\bf Abstract.} \Abstract
%\np
}

%%%%%%%%%%%%%%%%%%%%%%%%  AC.tex %%%%%%%%%%%%%%%%%%%%%%%%%%%%%%
%
%   These Macros provides automatic counters in PlainTeX for: 
%	      Formulae
%       Chapters
%       Sections
%       Subsections
%       Theorems, Propositions, ect.
%       
%
%
%   Fonts definitions
%
\font\ChapTitle=cmbx12% at 17pt
\font\SecTitle=cmbx12% at 12pt
\font\SubSecTitle=cmbx12% at 12pt
%
%
%   Indent definitions
%
\def\NormalSkip{\parskip=5pt\parindent=15pt\baselineskip=12pt\PageNumbers%
\leftskip=0cm\rightskip=0cm}

\def\PageNumbers{\footline={\hss\tenrm\folio\hss}}
\def\NoPageNumbers{\footline={}}

%
%
%   Skips definitions
%
\def\np{\vfill\eject}
\def\ss{\vskip 5pt}
\def\ms{\vskip 15pt}
\def\bs{\vskip 30pt}

\def\ni{\noindent}
%
%
%   Counters definitions
%
\newcount\CHAPTER		          %
\newcount\SECTION		          %
\newcount\SUBSECTION		       %
\newcount\FNUMBER     % Define \FNUMBER to be the counter for formulae
%
%
%   Counters initializations
%
\CHAPTER=0		          % Initialize \CHAPTER to 0
\SECTION=0		          % Initialize \SECTION to 0
\SUBSECTION=0		       % Initialize \SUBSECTION to 0
\FNUMBER=0		          % Initialize \FNUMBER to 0
%
%
% Definition of Plain style for chapters numbering
%
\long\def\NewChapter#1{\global\advance\CHAPTER by 1%
\np\NoPageNumbers\ \vfil\parindent=0cm\leftskip=2cm\rightskip=2cm{\ChapTitle #1\hfil}%
\vskip 4cm\ \vfil\eject\SECTION=0\SUBSECTION=0\FNUMBER=0\NormalSkip}
%
% 
% Definition of Plain style for sections numbering
%
\gdef\Mock{}
\long\def\NewSection#1{\bs\global\advance\SECTION by 1%
\ni{\SecTitle \ifnum\CHAPTER>0 \the\CHAPTER.\fi\the\SECTION.\ #1}\ms\SUBSECTION=0\FNUMBER=0}
\def\CurrentSection{\global\edef\Mock{\the\SECTION}} %
%
% Definition of Plain style for subsections numbering
%
\long\def\NewSubSection#1{\global\advance\SUBSECTION by 1%
{\SubSecTitle \ifnum\CHAPTER>0 \the\CHAPTER.\fi\the\SECTION.\the\SUBSECTION.\ #1}\ss\FNUMBER=0}
%
%
% Definition of Plain style for formulae numbering
%
%
\def\RightFormulaNumber{\global\advance\FNUMBER by 1\eqno{\fopen\the\FNUMBER\fclose}}
\def\LeftFormulaNumber{\global\advance\FNUMBER by 1\leqno{\fopen\the\FNUMBER\fclose}}
\def\RightFormulaLabel#1{\global\advance\FNUMBER by 1%
\eqno{\fopen\the\FNUMBER\fclose}\global\edef#1{\fopen\the\FNUMBER\fclose}}
\def\LeftFormulaLabel#1{\global\advance\FNUMBER by 1%
\leqno{\fopen\the\FNUMBER\fclose}\global\edef#1{\fopen\the\FNUMBER\fclose}}
%
%
% Definition of Composed style for formulae numbering
%
\def\HeadNumber{\ifnum\CHAPTER>0 \the\CHAPTER.\fi%
\ifnum\SECTION>0 \the\SECTION.\ifnum\SUBSECTION>0 \the\SUBSECTION.\fi\fi}
\def\ComposedRightFormulaNumber{\global\advance\FNUMBER by 1%
\eqno{\fopen\HeadNumber\the\FNUMBER\fclose}}
\def\ComposedLeftFormulaNumber{\global\advance\FNUMBER by 1%
\leqno{\fopen\HeadNumber\the\FNUMBER\fclose}}
\def\ComposedRightFormulaLabel#1{\global\advance\FNUMBER by 1%
\eqno{\fopen\HeadNumber\the\FNUMBER\fclose}%
\global\edef#1{\fopen\HeadNumber\the\FNUMBER\fclose}}
\def\ComposedLeftFormulaLabel#1{\global\advance\FNUMBER by 1%
\leqno{\fopen\HeadNumber\the\FNUMBER\fclose}%
\global\edef#1{\fopen\the\FNUMBER\fclose}}
%
%
%  Definition of Plain style for Theorem Numbering 
%
\def\TheoremNumber{\global\advance\FNUMBER by 1 \fopen\the\FNUMBER\fclose}
\def\TheoremLabel#1{\global\advance\FNUMBER by 1\fopen\the\FNUMBER\fclose%
\global\edef#1{\fopen\the\FNUMBER\fclose}}
%
%
%  Definition of Composed style for Theorem Numbering 
%
\def\ComposedTheoremNumber{\global\advance\FNUMBER by 1 \fopen\HeadNumber\the\FNUMBER\fclose}
\def\ComposedTheoremLabel#1{\global\advance\FNUMBER by 1\fopen\HeadNumber\the\FNUMBER\fclose%
\global\edef#1{\fopen\HeadNumber\the\FNUMBER\fclose}}
%
%
%     Style definition
%
\def\fopen{(}\def\fclose{)}                 % Define Formula Number Delimiters
\def\fn{\ComposedRightFormulaNumber}        % Define Formula Number Style
\def\fl{\ComposedRightFormulaLabel}         % Define Formula Label  Style
             % Define Theorem Number Style
              % Define Theorem Label  Style

\def\Compare#1#2{\message{^^J Compara \noexpand #1:=#1[#2]^^J}}

%%%%%%%%%%%%%%%%%%%%%%%%%%  BIBLIO.tex  %%%%%%%%%%%%%%%%%%%%%%%%%%%%%
%%%%%%%%
%
%  -It provides macros for references
%   A file with \bib{}{} declarations should be \input at the beginning
%   Then one can refer to them by \ref{}
%   References are stored in \refs{} (in order of quotation) and displayd by \Biblio
%
%       \bib{tokenname}{Title}
%       \ref{tokenname}
%       \refs
%       \Biblio
%       \LogRef{tokenname}
%
%       \eightpoint
%       \tenpoint
%
%%%%%%%%

\def\ni{\noindent}
\def\ss{\vskip 5pt}
\def\ms{\vskip 10pt}

\font\eightrm=cmr8

\def\noex{\noexpand}

\def\refs{}
\def\empty{\#}
\def\BibNumber{}
\def\BibTitle{}

\newcount\BNUM
\BNUM=0

\def\bib#1#2{\gdef#1{\global\def\BibNumber{\empty}\global\def\BibTitle{#2}}}

\def\ref#1{#1%%%%
\if\BibNumber\empty \global\advance\BNUM 1%%%%
\message{reference[\BibNumber]}\message{}%%%
\global\edef\refs{\refs \ss\ni[\the\BNUM]\ \BibTitle}%%%%
\global\edef#1{\noex\global\noex\edef\noex\BibNumber{[\the\BNUM]}%%%%
 \noex\global\noex\edef\noex\BibTitle{\BibTitle}}%%%%
{\bf [\the\BNUM]}%%%%
\else%%%%
{\bf \BibNumber}%%%%
\fi}

\def\Biblio{{\refs}}

%%%%%%%%%%%%%%%%%%%%%%%%%%%%%%%%%%%%%%%%%%%%%%%%%%%%%%%%%%%%%%%%%%%%%%%%%

%%%%%%%%%%%%%%%%%%%%%%%%%%%%  MACROS %%%%%%%%%%%%%%%%%%%%%%%%%%%%%

\def\Met{\hbox{\rm Met}}

\def\Diff{\hbox{\rm Diff}}
\def\Div{\hbox{\rm Div}}

\def\ds{\hbox{\bf ds}}
\def\d{\hbox{\rm d}}

\def\Aut{\hbox{\rm Aut}}
\def\dim{\hbox{\rm dim}}

\def\Lie{\hbox{\it \$}}
 
\def\calC{{\cal C}} 
\def\calE{{\cal E}} 
\def\calL{{\cal L}} 
\def\calW{{\cal W}} 
 
\def\calU{{\cal U}} 
\def\calB{{\cal B}}

%            Greek

\def\na{\nabla}
\def\la{\lambda}
\def\Si{\Sigma}
\def\si{\sigma}

\def\Om{\Omega}
\def\om{\omega}

\def\al{\alpha}
\def\be{\beta}
\def\Ga{\Gamma}

\def\de{\delta}

\def\te{\theta}

\def\R{{\Bbb R}}

%            Mat

\def\R{I \kern-.36em R}
\def\E{I \kern-.36em E}
\def\F{I \kern-.36em F}
\def\Co{I \kern-.66em C}
\def\id{1 \kern-.36em I}              %  Identita'

\def\del{\partial}                   %  Derivata parziale
                      %  Infinito
                   %  Intersezione
                  %  Unione disgiunta
                   %  Unione
            %  Valore assoluto

\def\QDE{{\offinterlineskip\lower1pt\hbox{\kern2pt\vrule width0.8pt%%%%
\vbox to8pt{\hbox to6pt{\leaders\hrule height0.8pt\hfill}\vfill%
\hbox to6pt{\hrulefill}}\vrule\kern3pt}}}
            %  aggiunta da Marco

%            Frecce

\def\arr{\rightarrow }            %  Freccia da applicazione
   %  Coinplicazione con indent
      %  Implicazione con indent
\def\QDE{\hbox{\ }\vrule height4pt width4pt depth0pt}                                                              %  Quadratino fine dimostrazione

\def\np{\vfill\eject}
\def\ni{\noindent}

\def\ss{\vskip 5pt}
\def\ms{\vskip 10pt}
\def\bs{\vskip 15pt}

%%%%%%%%%%%%%%%%%%%%%%%%%%%%%%   BIBLIO.def  %%%%%%%%%%%%%%%%%%%%%%%%%
\bib{\TaubNUTH}{C.\
J.\ Hunter, hep-th/9807010}

\bib{\TaubNUTHH}{S.\ W.\ Hawking, C.\ J.\ Hunter, hep-th/9808085}

\bib{\TaubNUTHHP}{ S.\ W.\ Hawking, C.\ J.\ Hunter, D.\ N.\ Page,
hep-th/9809035}

\bib{\HawHun}{ S.\ W.\ Hawking, C.\ J.\ Hunter, Class. Quantum Grav. {\bf 13}, (1996) 2735}

\bib{\HawHor}{ S.\ W.\ Hawking, G.\ T.\ Horowitz, Class. Quantum Grav. {\bf 13}, (1996)
1487}

\bib{\TaubNUTP}{ D.\ N.\ Page, Phys.\ Lett.\ {\bf 78B}, (1978) 249}

\bib{\Misner}{ C.\ W.\ Misner, J.\ Math.\ Phys.\ {\bf 4}, (1963) 924}

\bib{\WaldA}{V. Iyer and R. Wald, Phys. Rev. D {\bf 50},  (1994) 846;
R.M.\ Wald, J.\ Math.\ Phys., {\bf 31}, (1993) 2378}

\bib{\WaldB}{ I.\ Racz, R.\ M.\ Wald, Class. Quantum Grav. {\bf 9}, (1992) 2643}

\bib{\Remarks}{
L.\ Fatibene, M.\ Ferraris, M.\ Francaviglia, M.\ Raiteri, Ann.\ Phys.\ (in press),
hep-th/9810039;
L.\ Fatibene, M.\ Ferraris, M.\ Francaviglia, M.\ Raiteri, {\it Remarks on Conserved Quantities and
Entropy of BTZ Black Hole Solutions. Part I: the General Setting}, gr-qc/9902063;
L.\ Fatibene, M.\ Ferraris, M.\ Francaviglia, M.\ Raiteri, {\it Remarks on Conserved Quantities and
Entropy of BTZ Black Hole Solutions. Part II: BCEA Theory}, gr-qc/9902065
 }

\bib{\Kolar}{I.\ Kol{\'a}{\v r}, P.\ W.\ Michor, J.\ Slov{\'a}k, 
{\it Natural Operations in Differential Geometry}, 
(Springer--Verlag, New York, 1993)
}

\bib{\Saunder}{D.J.\ Saunders, {\it The Geometry of Jet Bundles},
Cambridge University Press, (Cambridge, 1989)
}

\bib{\Lagrange}{M. Ferraris, M. Francaviglia,
in: {\it Mechanics, Analysis and Geometry: 200 Years after Lagrange};
 M. Francaviglia ed., Elsevier Science Publishers B.V., (Amsterdam, 1991) 451}

\bib{\Ferraris}{M. Ferraris, in:
{\it Proceedings of the Conference on Differential Geometry and Its Applications},
Part 2, Geometrical Methods in Physics;
D.\ Krupka ed., (Brno, 1984) 61}

\bib{\Robutti}{M.\ Ferraris, M.\ Francaviglia and O.\ Robutti, in: {\it G\'eom\'etrie et Physique},
Proceedings of the {\it Journ\'ees Relativistes 1985} (Marseille, 1985); 
Y.\ Choquet-Bruhat, B.\ Coll, R.\ Kerner, A.\ Lichnerowicz eds., Hermann, (Paris, 1987) 112}

\bib{\Katz}{J.\ Katz, Class.\ Quantum Grav., {\bf 2}, (1985) 423}

\bib{\CADMC}{M.\ Ferraris and M.\ Francaviglia, Gen.\ Rel.\ Grav., {\bf 22}, (9), (1990) 965}

\bib{\BY}{J.\ D.\ Brown, J.\ W.\ York, Phys.\ Rev.\ D{\bf 47} (4), (1993) 1407;
J.\ D.\ Brown, J.\ W.\ York, Phys.\ Rev.\ D{\bf 47} (4), (1993) 1420}

\bib{\OurBY}{L. Fatibene, M. Ferraris, M. Francaviglia, M. Raiteri,
{\it  A Comparison between Quasi--Local Energy and Noether Charges}, (to appear)}

\bib{\GH}{G.\ W.\ Gibbons, S.\ W.\ Hawking, Phys.\ Rev.\ D{\bf 15} (10), (1977) 2752}

\bib{\RT}{T.\ Regge, C.\ Teitelboim, Annals of Physics {\bf 88}, (1974) 286.}

\bib{\AltriTaub}{
A.\ Chamblin, R.\ Emparan, C.\ V.\ Johnson, R.\ C.\ Myers, hep-th/9808177;
R.\ B.\ Mann, hep-th/9904148.}

%%%%%%%%%%%%%%%%%%%%%%%%%%%%%%  											  %%%%%%%%%%%%%%%%%%%%%%%%%

%%%%%%%%%%%%%%%%%%%%%%%%%%%%%%%%%%%%%%%

\def\titolo{The Entropy of Taub-Bolt Solution}
\def\autori{%
L.\ FATIBENE, %\note{E-mail: fatibene@dm.unito.it},
M.\ FERRARIS, %\note{E-mail: ferraris@dm.unito.it},
M.\ FRANCAVIGLIA, %\note{E-mail: francaviglia@dm.unito.it},
M.\ RAITERI.%\note{E-mail: raiteri@dm.unito.it}.
}

\def\indirizzo{Dipartimento di Matematica, Universit\`a degli Studi di Torino,\par
\ni Via Carlo Alberto 10, 10123 Torino, Italy}

\def\Abstract{{\eightrm
A geometrical framework for the definition of  entropy in General Relativity via N\"other theorem is
briefly recalled and the 
entropy of Taub-Bolt Euclidean solutions of Einstein equations is then obtained as an application.
The computed entropy agrees with previously known results, obtained by statistical
methods. It was generally believed that the entropy of Taub-Bolt solution
could not be computed via N\"other theorem, due to the particular structure of
singularities of this solution. We show here that this is not true.
The Misner string singularity is, in fact, considered and its contribution to the
entropy is analyzed. 
As a result, in our framework entropy does not obey the ``one--quarter area law" 
 and it is not directly related to horizons,  as sometimes erroneously suggested in current
literature on the subject.}}

%\vglue 103.3pt
\raggedbottom

\MainHead

\vglue 23.3pt
%\np
\pagenumbers

%%%%%%%%%%%%%%%%%%%%%%%%%%%%%%%%%%%%%%%%%%%%%%%%%%%%%%%%%%%%%%%%%%%%%%
%%%%%%%%%%%%%%%%%%%%%%%%%%%%%%%%%%%%%%%%%%%%%%%%%%%%%%%%%%%%%%%%%%%%%%
%%%%%%%%%%%%%%%%%%%%%%%%%%%%%%%%%%%%%%%%%%%%%%%%%%%%%%%%%%%%%%%%%%%%%%
%%%%%%%%%%%%%%%%%%%%%%%%%%%%%%%%%%%%%%%%%%%%%%%%%%%%%%%%%%%%%%%%%%%%%%
\NewSection{Introduction}

The Taub-Bolt and the Taub-NUT metrics are families of asymptotically locally flat (ALF) solutions
of Einstein field equations (without cosmological constant) in Euclidean (positive) signature.
We decided to deal with the Euclidean sector because in this case the entropy was already known in
literature by  the Euclidean path integral method and the Hamiltonian $(3+1)$ treatment of the action
functional (see \ref{\TaubNUTH}, \ref{\TaubNUTHH}\ and \ref{\TaubNUTHHP}).
In this way we know a priori which is the  result we are expected to produce. However, since our
framework applies without particular {\it a priori} requirements on signature or dimension also
the Lorentzian Taub-NUT metric seems to be susceptible of a similar analysis (though we do  not
investigate it here).

In these solutions the surfaces of constant radius $r$ have, at infinity, the topology of 
a non trivial  bundle $S^3\arr S^2$ with fiber  $S^1$ (of constant lenght), 
instead of the {\it usual} trivial topology $S^1\times S^2\arr S^2$  as 
in  asymptotically flat Euclidean solutions.
This non-trivial topology causes the first problem when attempting to derive
Taub-Bolt entropy, by trying to dimensionally reduce the action functional 
along the orbit of the vector $\del_\tau$, i.e. along the $U(1)$ isometry (see \ref{\TaubNUTH}).
Deeply related to this kind of problem is the fact that, as already noticed in the literature,
Taub-Bolt solution contains a Misner string (see \ref{\Misner}), i.e.\ a two dimensional surface
singularity running along the $z$-axis. This singularity is not enclosed by a Killing horizon and it
contributes to the entropy. For these reasons it was generally believed that the geometrical
approach (see
\ref{\WaldA}, \ref{\WaldB}\ and references quoted therein) based on the integration on  cross
sections of Killing horizons does not apply to this case. In a suitable sense our framework
encompasses and generalizes that ``older'' approach and shows that these conclusions do not hold.
This apparent paradox is due to the following facts: first, entropy is not directly related to
cross sections of Killing horizons;   secondly, there is no need of requiring horizons to be Killing
  since there is no need of extending them to bifurcate Killing horizons (see \ref{\Remarks}).

We shall here introduce Euclidean, asymptotically locally flat Taub-Bolt solutions together with the 
geometric notation used to compute their entropy.
Further details on Taub-Bolt and Taub-NUT solutions may be found in \ref{\TaubNUTH},
\ref{\Misner}, \ref{\AltriTaub}\ and \ref{\TaubNUTP}.
The entropy of Taub-Bolt solutions was already computed in \ref{\TaubNUTHH}\ and \ref{\TaubNUTHHP},
in the framework for conserved quantities due to Brown and York (see \ref{\BY}).
Our framework is instead based on a direct application of N\"other theorem to a covariant first order
Lagrangian for General Relativity (see \ref{\Remarks}, \ref{\CADMC}), but  it can be shown to be
deeply related to Brown and York framework. In a forthcoming paper of ours we shall investigate the
relations between these formalisms (see
\ref{\OurBY}). We shall not enter  here in details, but some similarity has to be noticed:
first, both methods compute the {\it relative conserved quantities and entropy with respect to a
fixed background}, regarding {\it absolute quantities} to be basically undefined;
second, both prescriptions reduce to ADM quantities (possibly corrected by ``Regge-Teitelboim
terms"; see \ref{\RT}) when the latter applies; third, both methods apply to a wider class
of situations than the standard ADM formalism for conserved quantities.
However, some differences   are to be stressed: first, our method is not restricted to Euclidean
signature; second, our analysis is completely covariant and basically independent on a spacetime
ADM foliation.

\NewSection{Notation}

We shall hereafter recall standard geometric notation (see \ref{\Saunder} and \ref{\Kolar})
as well as the  geometric framework for conserved quantities and entropy (\ref{\Remarks}).

Let $\calC=(C,M,\pi,F)$ be the {\it configuration bundle} of a field theory and $(x^\mu,y^i)$ a
system of fibered coordinates.
Let us denote by $J^k\calC$ the {\it $k^{th}$ order prolongation of $\calC$}, having
$(x^\mu,y^i,y^i_\mu,\dots,y^i_{\mu_1\dots\mu_k})$ as {\it natural fibered coordinates};
$y^i_{\mu_1\dots\mu_h}$ are understood to denote derivatives of local sections and are therefore meant
to be symmetric in their lower indices $(\mu_1\dots\mu_h)$.
If $\Xi=\xi^\mu\del_\mu+\xi^i\del_i$ is a projectable vector field over $\calC$,
one can define its {\it $k^{th}$ order prolongation}
$j^k\Xi=\xi^\mu\del_\mu+\xi^i\del_i+\xi^i_\mu\del_i^\mu+\dots+
\xi^i_{\mu_1\dots\mu_k}\del_i^{\mu_1\dots\mu_k}$ over $J^k\,\calC$ by setting
$$
\left\{
\eqalign{
&\xi^i_\mu=\d_\mu\xi^i-y^i_\nu\d_\mu\xi^\nu    \cr
&\dots\cr
&\xi^i_{\mu_1\dots\mu_k}=\d_{\mu_1}\xi^i_{\mu_2\dots\mu_k}-y^i_{\nu\mu_2\dots\mu_{k}}\d_{\mu_1}\xi^\nu   
\cr }\right.
\fl{\kProlongation}$$
where
$$
\d_\mu={\del\over \del x^\mu}+y^i_\mu {\del\over \del y^i}+y^i_{\mu\nu} {\del\over \del y^i_{\nu}}+
\dots+y^i_{\mu\nu_1\dots\nu_h} {\del\over \del y^i_{\nu_1\dots\nu_h}}+\dots
\fn$$
denotes the {\it total derivative operator} (at any order).

The bundle $\calC$ is assumed to be a {\it natural bundle}  (see \ref{\Kolar}): then there exists an
action of spacetime diffeomorphisms $\Diff(M)$ on $\calC$, i.e.\ a canonical group homomorphism
$\Diff(M)\arr\Aut(\calC)$ into the automorphisms of $\calC$.
Any natural bundle is associated to the $s$-frame bundle $L^s(M)$ for some finite and minimal
{\it order $s$} (see
\ref{\Kolar}).
Examples of natural bundles are the {\it tangent bundle $TM$} as well as any {\it tensor bundle
over $M$} (order $s=1$), or the {\it bundle of linear connections over $M$} (order $s=2$).
In particular, since we are interested in applications to General Relativity, the bundle
$\calC=\Met(M,\eta)$ of all metrics of signature $\eta=(p,q)$ is natural;
for $\eta=(n,0)$ we recover the Euclidean case, while for $\eta=(n-1,1)$ we recover the Lorentzian
case  ($n=\dim(M)$).
Being $\calC$ a natural bundle, any spacetime vector field $\xi$ canonically induces a vector field
$\hat \xi$ over $\calC$ which is called the {\it natural lift} of $\xi$ over $\calC$.
Clearly, $\hat\xi$ can be then prolonged to $J^k\calC$ by specializing equation
$\kProlongation$.

A {\it Lagrangian of order $k$} is a fibered morphism $L:J^k\calC\arr A_n(M)$
into the bundle $A_n(M)$ of $n$-forms over $M$ ($n=\dim(M)$).
Locally, one has
$$
L=\calL(x^\mu,y^i,y^i_\mu,\dots,y^i_{\mu_1\dots\mu_k})\>\ds
$$
where $\ds$ is the standard local volume of $M$  and $\calL$ is called the {\it
Lagrangian density}.
One can define the morphism
$\de L$ representing the {\it variation
of the Lagrangian}, i.e.\ for any vertical vector field $X$ over $\calC$
$$
\eqalign{
&\de L:J^k\calC\arr V^\ast(J^k\calC)\otimes A_n(M)\cr
&<\de L \circ j^k\si\>\vert\> j^kX>= \left[ {\d\over \d s}\left( L\circ J^k\Phi_s\circ
j^k\si\right)\right]_{s=0}\cr }
\fn$$
where $J^k\Phi_s$ is the (vertical) flow of $j^kX$.
We recall that $V^\ast(J^k\calC)$ is the dual of the {\it vertical bundle} $V(J^k\calC)$
where $j^kX$ lives, so that $<\de L  \>\vert\> j^kX>$
is a fibered morphism from $ J^k\calC$ into $ A_n(M)$. Once it is computed along the prolongation 
$j^k\si$ of any section $\si$ it produces an $n$-form over $M$. The standard {\it first variation
theorem} ensures the existence of a unique {\it Euler-Lagrange morphism} $\E(L):J^{2k}\calC\arr
V^\ast(\calC)\otimes A_{n}(M)$ and a family of {\it Poincar\'e-Cartan morphisms}
$\F(L,\Ga):J^{2k-1}\calC\arr V^\ast(J^{k-1}\calC)\otimes A_{n-1}(M)$ satisfying the 
{\it first vatiation formula}
$$
<\de L\>\vert\> j^kX>=<\E(L)\>\vert\> X>+\Div <\F(L,\Ga)\>\vert\> j^{k-1} X>
\fl{\FVF}$$
where $\Div$ denotes the {\it formal divergence operator} which acts on morphisms $\om:J^h\calC\arr
A_l(M)$ producing the morphisms $\Div\>\om:J^{h+1}\calC\arr A_{l+1}(M)$,
defined by the equation $(\Div\>\om)\circ j^{h+1}\si=\d(\om\circ j^{h}\si)$.
The  Poincar\'e-Cartan morphism is 
parametrized by a fibered
connection $\Ga$  (see
\ref{\Ferraris}, \ref{\Lagrange}); in  applications to  General Relativity it is standard to take
$\Ga$ to be  the Levi-Civita connection of the dynamical metric.

The Lagrangian $L$ is {\it natural} if it is generally covariant with respect to the action of
$\Diff(M)$ over $\calC$.
Infinitesimally, for any spacetime vector field $\xi$ and any section $\si$ of $\calC$,
the following identity must hold:
$$
<\de L \circ j^k\si\>\vert\> j^k\Lie_\xi\si>=\Lie_\xi (L\circ j^k\si)
\fl{\CondCov}$$ 
where
$$
\Lie_\xi\si= T\si(\xi)-\hat\xi\circ \si
\fl{\LieDerivativeOfSection}$$
is the {\it Lie derivative} of the section $\si$.

For standard General Relativity one can choose the Hilbert-Einstein Lagrangian
$$
L_{_{HE}}= {1\over 16\pi} \sqrt{g}\>g^{\mu\nu}\>R_{\mu\nu} \>\ds
\fl{\HEL}$$ 
where $R_{\mu\nu}$ denotes the Ricci tensor of the covariant metric $g^{\mu\nu}$
and $\sqrt{g}$ denotes the square root of the absolute value of the determinant of the metric.
This  is a natural Lagrangian.

Because of the covariance condition $\CondCov$ and the first variation formula
$\FVF$, for any spacetime vector field $\xi=\xi^\mu\>\del_\mu$ 
we can define the {\it N\"other current}
and the {\it work current} given respectively by:
$$
\eqalign{ 
&\calE(L,\xi)=<\F(L,\Ga)\>\vert\>j^{k-1}\Lie_\xi\si>-i_\xi \> L\cr
&\calW(L,\xi)=-<\E(L)\>\vert\>\Lie_\xi\si>\cr
}
\fn$$
Both the N\"other current
$\calE(L,\xi):J^{2k-1}\calC\arr A_{n-1}(M)$ and the work current
$\calW(L,\xi):J^{2k}\calC\arr A_{n}(M)$
are well-defined global bundle morphisms (provided   $\Lie_\xi\si$ is interpreted as the
{\it formal Lie derivative}, i.e.\ the bundle morphism $J^1\calC\arr V(\calC)$ which reduces to the Lie
derivative of a section $\LieDerivativeOfSection$ when evaluated along $\si$).
From $\CondCov$ and  $\FVF$
the following conservation law holds
$$
\Div\>\calE(L,\xi)=\calW(L,\xi)
\fn$$
which expresses {\it N\"other theorem}.
Notice that in this way N\"other theorem is expressed in terms of bundle
quantities; one can pull-back all currents along a section $\si$
of $\calC$ obtaining the N\"other theorem as a relation between forms over
$M$
$$
\eqalign{ 
&\calE(L,\xi,\si)=\calE(L,\xi)\circ j^{2k-1}\si\cr
&\calW(L,\xi,\si)=\calW(L,\xi)\circ j^{2k}\si\cr
}
\qquad\qquad
\d\>\calE(L,\xi,\si)=\calW(L,\xi,\si)
\fn$$
Whenever $\si$ is a solution of field equations, the work $\calW(L,\xi,\si)$ vanishes so that
$\calE(L,\xi,\si)$ is a closed form on $M$, i.e.\
$\calE(L,\xi)$ is conserved {\it on-shell}.

If  $\calC$ is a natural bundle of order $s$, i.e. it is associated to $L^s(M)$,
the current $\calE(L,\xi)$ (as well as $\calW(L,\xi)$) is a linear
combination of $\xi^\nu$ and its symmetrized covariant derivatives
(with respect to $\Ga$) up to the (finite)
order $k+s-1$ ($s$, respectively).
Integrating covariantly by parts one can {\it algorithmically} recast
the currents in the following form
$$
\eqalign{ 
&\calE(L,\xi)= \tilde\calE(L,\xi)+\Div\>\calU(L,\xi)\cr
&\calW(L,\xi)=\calB(L,\xi) +\Div\>\tilde\calE(L,\xi)\cr
}
\fn$$
The current $\calB(L,\xi)$ turns out to be identically zero
(which is known as {\it the generalized Bianchi identity}), while 
$\tilde\calE(L,\xi)$ is called the {\it reduced current}
(which vanishes when pulled-back along a solution $\si$ of field equations)
and $\calU(L,\xi)$ is called the {\it superpotential}.
Thence in any natural theory (as well as in the larger class
of {\it gauge-natural theories}; see \ref{\Kolar}, \ref{\Remarks})
the N\"other currents $\calE(L,\xi)$ are always exact {\it on-shell},
regardless of the topology of   spacetime $M$ (see \ref{\Robutti}).

Then a conserved quantity    associated to
$\xi$ and along a solution $\si$ is defined in a region $D\subset M$ as
$$
Q_D(L,\xi,\si)=\int_D \calE(L,\xi,\si)=\int_{\del D} \calU(L,\xi,\si)
\fl{\CCQ}$$
In applications to General Relativity only {\it conserved quantities
relative to some background} $\bar g$ are in general defined (see \ref{\TaubNUTH},
\ref{\Remarks}, \ref{\BY} and references quoted therein). For the metric field $g$ and the background
$\bar g$ one can define the following {\it covariant} (i.e.\ natural) Lagrangian
$$
\eqalign{
L_{_{\hbox{\sevenrm Tot}}}=&L_{g} +L_{g\bar g}+L_{\bar g}=\cr
=& {1\over 16\pi}\Big[
\sqrt{g}\>g^{\mu\nu}\>R_{\mu\nu} 
-d_\la (\sqrt{g}\>g^{\mu\nu} w^\la_{\mu\nu})
-\sqrt{\bar g}\>\bar g^{\mu\nu}\>\bar R_{\mu\nu} 
\Big]\>\ds
\cr}
\fl{\FOL}$$
where we set $R_{\mu\nu}$ and $\bar R_{\mu\nu}$ for the Ricci tensor
of the metric $g$ and of the background $\bar g$, respectively, and
$$
\eqalign{
&\Ga^\la_{\mu\nu}\quad \hbox{Levi-Civita connection of $g$}\cr
&\bar\Ga^\la_{\mu\nu}\quad \hbox{Levi-Civita connection of $\bar g$}\cr
&\cr
}
\qquad\qquad
\eqalign{
&u^\la_{\mu\nu}=\Ga^\la_{\mu\nu}- \de^\la_{(\mu} \Ga^\al_{\nu)\al}\cr
&\bar u^\la_{\mu\nu}=\bar \Ga^\la_{\mu\nu}
   -\de^\la_{(\mu} \bar\Ga^\al_{\nu)\al}\cr
&w^\la_{\mu\nu}= u^\la_{\mu\nu}-\bar u^\la_{\mu\nu}\cr
}
\fn$$

By defining the tensor quantities $q^\la_{\mu\nu}= \Ga^\la_{\mu\nu}-\bar \Ga^\la_{\mu\nu}$,
the Lagrangian $\FOL$ can be recasted as
$$
L_{_{\hbox{\sevenrm Tot}}}
={1\over 16\pi}\Big[
\sqrt{g}\>g^{\mu\nu}\>\big( q^\rho_{\mu\si}q^\si_{\rho\nu}-q^\si_{\si\rho}q^\rho_{\mu\nu}\big)
+ \big(\sqrt{g}\>g^{\mu\nu}-\sqrt{\bar g}\>\bar g^{\mu\nu}\big)\>\bar R_{\mu\nu}
\Big]\>\ds
\fn$$
showing that it is first order in $g$ and second order in $\bar g$.
We also remark that the Lagrangian $L_{_{\hbox{\sevenrm Tot}}}$ vanishes when computed on the background
$g=\bar g$.
A forthcoming paper (see \ref{\OurBY}) will be devoted to a detailed discussion about the choice of
the Lagrangian
$\FOL$ and the comparison with other choices of the action functional suited to deal with conserved
quantities relative to a background  (see also \ref{\BY}, \ref{\HawHor},\ref{\HawHun}\ and references quoted
therein).

The field equations induced by the Lagrangian $\FOL$ are vacuum Einstein equations for $g$ and
$\bar g$, respectively. The Poincar\'e-Cartan morphism is defined by
$$
\eqalign{
<\F(L_{_{\hbox{\sevenrm Tot}}})\>\vert&\>  j^1\Lie_\xi \si>=<\F(L_{g})\>\vert\>  j^1\Lie_\xi g>
+<\F(L_{g\bar g})\>\vert\>  j^1\Lie_\xi \si>+\cr
&\qquad\qquad\quad
+<\F(L_{\bar g})\>\vert\>  j^1\Lie_\xi \bar g>=\cr
=&{1\over 16\pi}\Big[
\sqrt{g}\big( g^{\la\al}g_{\mu\nu}-\de^\la_{(\mu}\de^\al_{\nu)}\big)
    \na_\al (\Lie_\xi g^{\mu\nu})
-\Lie_\xi\big( \sqrt{g}\> g^{\mu\nu} \> w^\la_{\mu\nu}\big)+\cr
&\qquad\qquad\quad
-\sqrt{\bar g}\big( \bar g^{\la\al}\bar g_{\mu\nu}-\de^\la_{(\mu}\de^\al_{\nu)}\big)
    \bar \na_\al(\Lie_\xi\bar g^{\mu\nu})
\Big]\>\ds_{\la}\cr
}
\fn$$
where $\si$ denotes here the pair $(g,\bar g)$ while
  $\na_\al$ and $\bar \na_\al$ denote the covariant derivatives induced by
$\Ga^\al_{\be\mu}$ and $\bar \Ga^\al_{\be\mu}$, respectively.

The Lagrangian $\FOL$ is natural (when regarded as a Lagrangian for the
field pair $(g,\bar g)$) and thence allows a superpotential
$$
\eqalign{
\calU(L_{_{\hbox{\sevenrm Tot}}}, \xi)=&\>\calU(L_{g}, \xi)+\calU(L_{g\bar g}, \xi)+
\calU(L_{\bar g}, \xi)=\cr
=&{1\over 16\pi}\Big[
\sqrt{g}\>\na^\be \xi^\al + \sqrt{g}\> g^{\mu\nu} w^\be_{\mu\nu} \xi^\al
-\sqrt{\bar g}\>\bar \na^\be \xi^\al
\Big]\>\ds_{\al\be}\cr
}
\fl{\Superpotenzial}$$

If one considers an ADM foliation of spacetime $M$ by means of
{\it ``spacelike''} hypersurfaces $\Si_\tau=\{\tau=\hbox{\it const}\}$ and
denotes by $\infty$ the spatial infinity of each leaf $\Si_\tau$,
the {\it total conserved quantity of $g$ relative to $\bar g$} with respect to a spacetime vector
field $\xi$ is defined according to $\CCQ$ by
$$
Q_{_{\hbox{\sevenrm Tot}}}=\int_\infty \calU(L_{_{\hbox{\sevenrm Tot}}},\xi)\circ j^{1}\si
\fl{\ConservedQuantity}$$

The variation of the conserved quantity $Q_{_{\hbox{\sevenrm Tot}}}$
along a (vertical) vector field $X$ is given by
$$
\eqalign{
 \de Q_{\hbox{\sevenrm Tot}}=&
\int_\infty\Big(
\de \calU(L_{g},\xi)
-i_\xi<\F(L_{g})\>\vert\> j^{1}X>
\Big)\circ j^1g\,\,+\cr
&+\int_\infty\Big(
\de \calU(L_{\bar g},\xi)
-i_\xi<\F(L_{\bar g})\>\vert\> j^{1}X> 
\Big)\circ j^1\bar g
\cr}
\fl{\VarCQ}$$
provided the background $\bar g$ is chosen to approach the metric $g$ at infinity
(a choice which of course selects a class of suitable backgrounds) so that the following holds true
$$
\de \calU(L_{g\bar g}, \xi)\Big\vert_\infty= -i_\xi<\F(L_{g})\>\vert\> j^{1}X>\Big\vert_\infty
-i_\xi<\F(L_{\bar g})\>\vert\> j^{1}X> \Big\vert_\infty
\fn$$

The integral of the quantity
$\de \calU(L_{_{\hbox{\sevenrm Tot}}},\xi)
-i_\xi<\F(L_{_{\hbox{\sevenrm Tot}}})\>\vert\> j^{1}X>$   equals $\de Q_{\hbox{\sevenrm Tot}}$
as given by $\VarCQ$, since  for any Lagrangian $L$  the quantity
$\de \calU(L_{},\xi ) -i_\xi<\F(L_{} )\>\vert\> j^{k-1}X>$
is independent on the addition of pure divergence terms to the Lagrangian.
Thence it is a good candidate for {\it the density of variation of conserved
quantities}.
Notice also that   ``absolute conserved quantities"
may (and in some examples they in fact do) diverge, so that only {\it relative}
quantities are defined in general (as we already mentioned above).

This geometric, global and variational framework for conserved quantities
is particularly suited to deal with black hole entropy.
Entropy is a state function $S$ which satisfies the (generalised) first principle of
thermodynamics
$$
\de m = T\de S + \Om\de j
\fl{\FPT}$$
where $m$ is the {\it total mass}, $j$ is the {\it total angular momentum},
$\Om$ is the {\it angular velocity of the black hole horizon} and $T$ is the
{\it temperature} of the Hawking radiation irradiated by the black hole.
Further work terms may occur in general on the r.h.s.\ of equation $\FPT$
due to gauge symmetries;
they are not considered here since we are going to apply this framework
to a natural theory (see \ref{\Remarks}).

Inverting $\FPT$ one immediately obtains for the variation of the entropy
$$
\de S={1\over T}\Big(\de m-\Om \de j\Big)
\fl{\VEa}$$
The quantity $(\de m-\Om \de j)$ on the r.h.s.\ is readily recognised as the
N\"other conserved quantity associated to $\xi=\del_\tau+\Om\del_\phi$, so
that equation $\VEa$ can be recasted as
$$
\de S={1\over T} \int_\infty \Big(
\de \calU(L_{_{\hbox{\sevenrm Tot}}},\xi) -i_\xi<\F(L_{_{\hbox{\sevenrm Tot}}})\>\vert\>
j^{1}X>
\Big)\circ j^1\si
\fl{\VEb}$$
Quantities on the r.h.s.\ of this latter equation are known; equation
$\VEb$ can thence be regarded as a variational equation for $S$ that, if integrable,
can be integrated to provide at once the entropy and the
{\it ``first principle of thermodynamics''} $\FPT$ it obeys by its own definition.

The following theorem holds in any natural (as well as gauge-natural, see \ref{\Remarks})
theory.

\ms
{\sl
For any natural (as well as gauge-natural) Lagrangian $L$
let $\si$ be  a solution of field equations, $X$   a solution of linearized
field equations (i.e.\ it drags solutions into solutions) and
$\xi$ s a Killing vector for $\si$ (i.e.\ $\Lie_\xi\si=0$).
Then the form
$(\de \calU(L_{},\xi) -i_\xi<\F(L_{})\>\vert\> j^{k-1}X>)\circ j^{2k-1}\si$
is  closed.}

\ms
Specializing this theorem to both terms in the r.h.s. of $\VarCQ$,
we see that the integral in $\VEb$ does not depend on
the integration region $\infty$ but just on its homology. Let then
$\Si$ be any $(n-2)$-surface such that $\infty-\Si$ is a  homological
boundary; we can express the variation of 
entropy as
$$
\de S={1\over T} \int_\Si \Big(
\de \calU(L_{_{\hbox{\sevenrm Tot}}},\xi) -i_\xi<\F(L_{_{\hbox{\sevenrm Tot}}})\>\vert\>
j^{k-1}X>
\Big)\circ j^1\si
\fl{\VEc}$$
When dealing with black hole solutions this formula reproduces the ``one quarter area law" (see
\ref{\GH}).

We stress that in the definition $\VEc$ entropy  is not {\it a priori} related to spatial infinity
(as conserved quantities are), nor to the horizon of black hole (as sometimes it is asserted in the
literature). Due to this fact we may apply formula  $\VEc$ also to solutions other then black hole
ones.

\NewSection{Taub-Bolt Solution}

We are going to describe Taub-Bolt (and Taub-NUT) solutions of vacuum Einstein equations
in four-dimensional Euclidean General Relativity . We also give here a short summary of
the results obtained in \ref{\TaubNUTH}, \ref{\TaubNUTHH} and \ref{\TaubNUTHHP}, in order to
compare them with our computations in the next Section.

The Taub-Bolt metric
$$
\left\{
\eqalign{
&g_{_{TB}}=V_{_{TB}}\left( d\tau +2 N \cos\theta d\phi 
\right)^2+{dr^2\over{V_{_{TB}}}}+(r^2- N^2)(d\theta^2 +\sin^2\theta d\phi^2) \cr
&V_{_{TB}}={(r-2N)(r-N/2)\over{r^2-N^2}}\qquad   N \hbox{ constant }\cr
}\right.
\fl{\TB}$$
is regular if we assume $r> 2 N$ and we fix the period $\be$ of the Euclidean time
$\tau$ equal to $8\pi N$. This vacuum solution is not asymptotically flat but rather
{\it asymptotically locally flat} (ALF). The boundary at large $r$ is a $S^1$ bundle with
constant circumference $\be$ over a sphere $S^2$ parametrized by $\theta$ and $\phi$.
The fixed points set of the $U(1)$ isometry  $\tau$ is the two dimensional surface
$r=2N$ parametrized by $\theta$ and $\phi$. It is called a {\it bolt} and its area is equal to
$12\pi N^2$. Moreover, the metric $\TB$ has  NUT charge due to the presence of a Misner
string (which  is a two dimensional coordinate singularity) running along the $z$-axis
from the bolt out to infinity (see \ref{\Misner}). 

The first implication  of the asymptotic
behaviour  of Taub-Bolt metric is that it cannot be matched at infinity to flat space.
In order to deal with non-compact spacetimes one nevertheless needs a background
solution with the same asymptotic behaviour of the solution itself. In Taub-Bolt metric
the background is chosen to be the   self--dual Taub-NUT solution (see \ref{\TaubNUTH}). The general
Taub-NUT solution is:
$$
\left\{
\eqalign{
&g_{_{TN}}=V_{_{TN}}\left( d\tau +2 N \cos\theta d\phi 
\right)^2+{dr^2\over{V_{_{TN}}}}+(r^2- N^2)(d\theta^2 +\sin^2\theta d\phi^2) \cr
&V_{_{TN}}={r- N \over{r+ N }} \cr
}
\right.
\fl{\TNO}$$
It is regular if we assume $r> N$ and it has the same ALF behaviour of
Taub-Bolt metric. The fixed points set of the $U(1)$ isometry is now described by the
zeros of $V_{_{TN}}$. In this case however, the set $r=  N$, called the {\it nut},
 is   $0$ dimensional  because it occurs precisely when the  $2$-sphere
(parametrized by $(\theta,\phi)$) degenerates.  Again, there is a Misner string which now 
runs along the $z$  axis from the
nut out to infinity. 

We remark that both Taub-Bolt and Taub-NUT Euclidean solutions are obtained by Wick
rotating   the same Lorentzian Taub-NUT metric:
$$\left\{
\eqalign{
&g=-f(r)  \left( dt +2 l \cos\theta d\phi 
\right)^2+{dr^2\over{f(r) }}+(r^2+ l^2)(d\theta^2 +\sin^2\theta d\phi^2) \cr
&f(r)={ r^2-2 m r-l^2  \over{r^2+l^2 }} \qquad m,l \hbox{ constants}\cr
}
\right.
\fl{\LTN}$$
In order to avoid conical singularity at the origin, in the Euclidean sector one has to
fix the constant $m$ in terms of $l\rightarrow i\,N$. The two different fixings:
$$
m={5\over{ 4}} N
\qquad\qquad
m=   N  
\fl{\EuclMass}$$
give  the Taub-Bolt and the Taub-NUT Euclidean solutions, respectively (see  \ref{\TaubNUTP}).

In order to consider Taub-NUT metric as a well defined  background for Taub-Bolt spacetime one
requires  it to match  (at the suitable  asymptotic order) the Taub-Bolt solution $\TB$ on the
``large'' sphere of radius $\rho$ (see
\ref{\TaubNUTH}). To achieve this task,
it is necessary to rescale the imaginary time and the nut charge in $\TNO$. The result is the
matched Taub-NUT metric:
$$
\left\{
\eqalign{
&\bar g  =\mu^2 \bar V \left( d\tau +2 N \cos\theta d\phi 
\right)^2+{dr^2\over{\bar V }}+(r^2-\mu^2 N^2)(d\theta^2 +\sin^2\theta d\phi^2) \cr
&\bar V  ={  r-\mu N  \over{r+\mu N }}\qquad \mu=\left(1-{N\over{4\rho}}\right)\cr
}
\right.
\fl{\TN}$$
which agrees  with $\TB$ at order $O(1/\rho)$ on$r=\rho$.

The starting point for the analysis developed in  \ref{\TaubNUTH} and
\ref{\TaubNUTHH} is the action functional:
$$
I=-{1\over {16\pi}}\int_D d^4 x\sqrt{g}\, R(g)-{1\over
{8\pi}}\int_{\partial{D}} d^3x \,\sqrt{b}\left[\Theta(b)-\Theta(\bar b)\right]
\fl{\BYAzione}$$
where ${D}$ is a compact region of spacetime with regular boundary $\partial{D}$, namely
the $3$-sphere of radius $\rho$ which is let then go to infinity;
$b_{\mu\nu}$ is the metric induced on $\partial{D}$ and $\Theta(b)$ is the trace of
the extrinsic curvature of $\partial{D}$ in $ {D}$. The bar is again  used in order to
indicate the objects built out from the background metric.  

We recall that the volume
integral in $\BYAzione$, when evaluated along solutions, gives no contribution 
to the computation of the action. Moreover Taub-NUT metric $\TN$ matches
correctly the Taub-Bolt metric $\TB$ on
$\partial{D}$ so that the total action has a finite value:
$$
I=\pi N^2
\fl{\TotalAction}$$

We also remark that the action $\BYAzione$ is a possible choice for an   action
suited  to describe a manifold with a   boundary of fixed intrinsic
geometry (additional terms are needed if we are in presence of non orthogonal
boundaries, see \ref{\HawHun}). Roughly speaking it has a well defined variational principle if
subjected to these boundary conditions (i.e. fixed intrinsic   three--metric on the
boundary). Moreover $\BYAzione$ is defined in order to be identically zero when computed  along the
background solution $\bar g$.

We remark that another choice of a suitable action, instead of $\BYAzione$, is the action induced
by  the Lagrangian $\FOL$
$$
I_{_{\hbox{\sevenrm Tot}}}=\int_D L_{_{\hbox{\sevenrm Tot}}} \circ j^2\si 
\fl{\OurAzione}$$
Notice that if one computes the value of the total action $\OurAzione$
for the Taub-Bolt metric $\TB$ with respect to the Taub-NUT matched background $\TN$
one obtains $I_{_{\hbox{\sevenrm Tot}}}= \pi N^2$ as well.
Notice also that the same result is obtained using the Taub-NUT   solution $\TNO$ as a
background. These results are due to the fact that, fixing $\del D$ to be the $3$-sphere of
constant radius
$r=\rho$, the action $\BYAzione$ agrees with the action $\OurAzione$ as $\rho$ approaches infinity
(see \ref{\OurBY} for greater details).

We also recall (see \ref{\TaubNUTH}) the expression of the action in terms of the {\it Komar mass}:
$$
I={\be\over {2}}\, M_{\hbox{\sevenrm Kom}}
\fl{\MassaKomar}
$$
which, due to  $\TotalAction$, gives
$$
M_{\hbox{\sevenrm Kom}}={N\over{4}}
\fl{\MassaKomarNum}
$$
This result, again, must be viewed as relative to the background.

In order to calculate the entropy, a $3+1$ decomposition of the action $\BYAzione$ was performed
in \ref{\TaubNUTHH}.
The main result    is that entropy   is not just a
quarter of the area of the fixed points set  of the    Euclidean time
translation Killing vector field, i.e.  of the bolt (as it is for black hole solutions). A further
contribution is in fact due to the presence of the Misner string. According  to \ref{\TaubNUTHH} one
can associate the entropy of a solution to the obstructions in foliating   topologically
non trivial Euclidean spacetimes with
$\tau=const$  surfaces which do not intersect and which agree with Euclidean time at
infinity. These obstructions arise from the  presence of bolts as well as from Misner
strings. The  
$\tau=const$ surfaces have a boundary at the fixed points set and   around Misner
string  plus the usual boundary at infinity.  Additional boundaries are responsable  of
further contributions in evaluating the  formula for entropy. The final  formula obtained 
in \ref{\TaubNUTHH}\ and \ref{\TaubNUTHHP} is
$$
S={1\over{4}}\left( A_{\hbox{\sevenrm Bolt} }+\Delta A_{MS}\right)-\be H_{MS}
\fl{\HHEntropy}$$
where $A_{\hbox{\sevenrm Bolt} }=12\pi N^2$ is the area of the bolt,
$\Delta A_{MS}=-12\pi N^2 $ is
the difference between the area of the Misner string in the Taub-Bolt metric minus the same
area in the Taub-NUT solution. The term $ H_{MS}$ is the Hamiltonian surface term
evaluated on the boundary of the Misner string on the $\tau=const$ surfaces. It is given
by the shift of the $3+1$ decomposition of the metric times a component of the second
fundamental form of the constant $\tau$ surface. The difference 
of this term  from the background  is finite and it has been  computed to be 
$ H_{MS}=-N/8$ (see \ref{\TaubNUTHH}). Thus the entropy turns finally out to be:
$$
S=\pi N^ 2
\fl{\ValoreEntropia}$$

\NewSection{Conserved Quantities and Entropy}

Let us consider the Taub-Bolt metric $\TB$ together with the Taub-NUT matched solution $\TN$
as background. We shall hereafter specialize to these metrics the formulae introduced in Section
$2$.

The total conserved quantity $\ConservedQuantity$ for the superpotential $\Superpotenzial$
and the vector $\xi=\del_\tau$, can be readily computed as
$$
Q_{_{\hbox{\sevenrm Tot}}}={5\over 8} N +{N\over 8} - {N\over 2}= {N\over 4}
\fn$$
which agrees with expression $\MassaKomarNum$.
We remark that the pure divergence term $L_{g\bar g}$ in the Lagrangian $\FOL$
reverberates in conserved quantities by curing the {\it anomalous factor} (see \ref{\Katz});
also in this example, in fact, the conserved quantity associated to each Lagrangian
$L_{g}$ and $L_{\bar g}$ would equal one half of the expected value. 

Analogously, one can apply expression $\VarCQ$ to determine the variation of the conserved quantity
$$
\de Q_{_{\hbox{\sevenrm Tot}}}={5\over 4} \de N - \de N={1\over 4}\de N
\fn$$
Notice that the contributions due to the Taub-Bolt metric $\TB$ and the Taub-NUT matched solution
$\TN$ can be isolated reproducing the expected values (see expressions $\EuclMass$).
Notice also that pure divergence terms in the Lagrangian, as already remarked,
do not contribute to the {\it variation of conserved quantity}. 

According to the previous Section, let us consider ${1/ T}=\be=8\pi N$ in expression $\VEb$;
one easily obtains
$$
\de S= {1\over T} \de Q_{_{\hbox{\sevenrm Tot}}}= 2\pi N \de N
\fl{\OurVarEntropia}$$
which can be integrated to obtain $S=\pi N^2$. It again agrees with the expected value
for the total entropy $\ValoreEntropia$.

One can   check that the $2$-forms
$\left(\de \calU(L,\xi) -i_\xi<\F(L)\>\vert\> j^{k-1}X>\right) \circ j^1\si$ in $\VarCQ$
are separately closed for both the Lagrangians $L=L_{g}$ and  $L=L_{\bar g}$, as it follows also
from the general theory since
$\xi=\del_\tau$ is a Killing vector of both the Taub-Bolt metric and the Taub-NUT background.
Then the variation of the entropy can be re-written as an integral over a finite $2$-region $\Si$
homologous to $\infty$.
Of course, because of the Misner string, the region $\Si$ cannot be the Bolt
$S^2_{_{B}}=\{t=t_0,\> r=2N\}$ alone; to obtain a region homologous to
$\infty$ we must also consider two cones $C_{_{B}}=\{t=t_0,\> \te=\te_0\}$ wrapping around the
singularity (due to Misner string) for $\te_0\simeq 0$ and $\te_0\simeq \pi$.
Analogously for the Taub-NUT background, we can consider any region enveloping the nut surface
$S^2_{_{N}}=\{t=t_0,\> r\arr\mu N\}$ together with the cones $C_{_{N}}$ running now from the nut out  to infinity.
In this way one can compute various contributions to $\VEc$;  letting  then the cones close over the
$z$-axis   one obtains
$$
\eqalign{
&\de S_{_{B}}={1\over T}\int_{S^2_{_{B}}}\de \calU(L_{g},\xi) -i_\xi<\F(L_{g})\>\vert\>
j^{k-1}X>= 6\pi N\de N \cr
&\de S_{_{N}}={1\over T}\int_{S^2_{_{N}}}(\de \calU(L_{\bar g},\xi) -i_\xi<\F(L_{\bar g})\>\vert\>
j^{k-1}X>= -2\pi N\de N \cr
&\de S_{_{MS}}={1\over T}\int_{C_{_{B}}}\de \calU(L_{g},\xi) -i_\xi<\F(L_{g})\>\vert\> j^{k-1}X>=
4\pi N\de N \cr
&\de S_{_{\overline{MS}}}={1\over T}\int_{C_{_{N}}}\de \calU(L_{\bar g},\xi) -i_\xi<\F(L_{\bar
g})\>\vert\> j^{k-1}X>= 10\pi N\de N \cr
}
\fl{\Contributions}$$
Of course the total variation of the entropy
$$
\de S=\de S_{_{B}}+\de S_{_{MS}}-\de S_{_{N}}-\de S_{_{\overline{MS}}}= 2\pi N\de N
\fn $$
agrees with the established result
$\OurVarEntropia$ and can be integrated to agree again with expression $\ValoreEntropia$.
We remark that one can isolate the contribution due to the bolt $S_{_{B}}=3\pi N^2$,
which equals ${1\over 4} A_{\hbox{\sevenrm Bolt} }=3\pi N^2$.
We also remark that the difference of the contributions along the Misner string
$S_{_{MS}}-S_{_{\overline{MS}}}=-3\pi N^2$ equals one quarter of the difference of the areas of the
Misner string $\Delta A_{MS}=-3\pi N^2$ with respect of Taub-Bolt and Taub-NUT matched solutions.

\NewSection{Conclusions and Perspectives}

We reproduced the value of entropy of Taub-Bolt metric {\it relative} to the Taub-NUT matched
background by the geometric framework introduced in Section $2$.
In that framework entropy is not related to any horizon; in particular we do not need horizons to be
Killing so that the Misner string can be treated as well, re-producing the deviation from the 
one-quarter-area paradigm   already noticed by Hawking, Hunter and Page (see
\ref{\TaubNUTHH}, \ref{\TaubNUTHHP}).
Consequently, our results perfectly agree with the conclusions there achieved:
entropy is related to the obstruction to foliate the spacetime with a family of hypersurfaces
of costant $\tau$. In fact, the breakdown to foliation comes from coordinate singularities of the
metric (fixed points set, as the bolt and nut, as well as the Misner string).
These singularities are related to boundary terms in $\HHEntropy$ and $\Contributions$
which both depend on the $2$-homology of hypersurfaces $\Si_\tau$ at constant $\tau$.

We finally remark that calculations may be performed also by using the Taub-NUT metric $\TNO$ as
background in place of
$\TN$. The metric $\TNO$ agrees with the Taub-Bolt metric $\TB$ at infinity as well, but in this
case the matching condition required in \ref{\TaubNUTH} is not satisfied.
In our framework this additional matching condition is not necessary; in fact, all results
presented in Section $4$ are unalterated by setting $\TNO$ as a background.  

Future investigations will be devoted to the study of the Lorentzian case as well as 
the asymptotically locally anti-deSitter case (i.e.\ with cosmological constant).
Both these examples appear to fall into our scope since no hypotheses restrict dimension of the
spacetime, signature, or asymptotic behaviour of the metric.

\NewSection{Acknowledgments}

We are grateful to  M.\ Caldarelli, D.\ Klemm and V.\ Moretti of the
University of Trento for having drawn our attention to Taub-Bolt solution and its entropy, as well
as to C.\ J.\ Hunter for useful enlightenings on the geometrical structure of the Taub-Bolt
solution.

\NewSection{References}

\Biblio

\end